\documentclass[a4paper]{jpconf}
\usepackage{amsmath}
\usepackage{graphicx}
\usepackage{hyperref}
\usepackage{multirow}
\usepackage{caption}
\usepackage{makecell}

\begin{document}
\title{Automatic bony structure segmentation and curvature estimation on ultrasound cervical spine images - a feasibility study}

\author{Songhan Ge$^1$, Haoyuan Tian$^1$, Wei Zhang$^1$, Rui Zheng$^{1,2,*}$}

\address{$^1$School of Information Science and Technology, ShanghaiTech University, 201210, Shanghai, China}
\address{$^2$Shanghai Engineering Research Center of Energy Efficient and Custom AI, Shanghai, 201210, Shanghai, China}

\ead{$^{*}$zhengrui@shanghaitech.edu.cn}

\begin{abstract}

The loss of cervical lordosis is a common degenerative disorder known to be associated with abnormal 
spinal alignment. In recent years, ultrasound (US) imaging has been widely applied in the assessment of 
spine deformity and has shown promising results. The objectives of this study are to automatically segment 
bony structures from the 3D US cervical spine image volume and to assess the cervical lordosis on the key sagittal frames. 
In this study, a portable ultrasound imaging system was applied to acquire cervical spine image volume. 
The nnU-Net was trained on to segment bony structures on the transverse images and validated by 5-fold-cross-validation. 
The volume data were reconstructed from the segmented image series. An energy function indicating intensity levels and 
integrity of bony structures was designed to extract the proxy key sagittal frames on both left and right sides for the 
cervical curve measurement. The mean absolute difference (MAD), standard deviation (SD) and correlation between 
the spine curvatures of the left and right sides were calculated for quantitative evaluation of the proposed method. 
The DSC value of the nnU-Net model in segmenting ROI was 0.973. For the measurement of 22 lamina curve angles, 
the $MAD\pm SD$ and correlation between the left and right sides of the cervical spine were $3.591\pm3.432^\circ$ and 0.926, 
respectively. The results indicate that our method has a high accuracy and reliability in the automatic segmentation of the 
cervical spine and shows the potential of diagnosing the loss of cervical lordosis using the 3D ultrasound imaging technique.
    
\end{abstract}

\section{Introduction}
The cervical spine consists of seven cervical vertebrae (C1-C7). It acts as a critical structure that bridges the head and thoracic spine. 
Each vertebra includes a  vertebral body at the front and a vertebral arch at the back which surrounds the spinal cord~\cite{kaiser2023a}.
Cervical lordosis is the natural inward curvature of the neck that is essential for head balance and minimizing neck strain~\cite{gm2018d}. 
Studies show that the annual incidence of cervical spine pain ranges from 10.4\% to 
 21.3\%~\cite{nori2022}. Deviations from normal cervical lordosis, such as loss of cervical lordosis
and cervical hyperlordosis, can cause neck pain, headaches, and other problems~\cite{delen2023}. 
The diagnosis of cervical spine curvature can be performed using various imaging modalities, including X-ray, 
MRI (Magnetic Resonance Imaging), and CT (Computed Tomography). 
The Cobb angle measured on the standing radiograph is the gold standard for estimating cervical lordosis, 
and it is the angle formed by lines drawn from the lower end plates of the C2 and C7 vertebra~\cite{gore1986a}. 
Other methods for estimating cervical lordosis include the Jackson physiological stress line, 
the Harrison posterior tangent method the Chin-Brow Vertical Angle (CBVA) and C2-C7 Sagittal Vertical Axis 
(SVA)~\cite{jackson1994,jackson2010a,suk2003,harrison2000a}, which are as well measured on radiographs and 
provide comprehensive curvature evaluation from different perspectives. 
However, X-ray and CT scans are not suitable for frequent diagnostic use because of the radiation exposure, 
and MRI  is costly and not advisable for patients with metal implants because of its intense magnetic field.

 Ultrasound imaging is a non-ionizing, portable, and cost-effective imaging technique. 
 It provides real-time imaging capabilities, rendering it a dependable tool for 
 evaluating musculoskeletal structures in clinical assessments.
 Recent research has demonstrated that ultrasound imaging is a reliable technique for 
 measuring spinal curvature, such as scoliosis~\cite{zheng2016}.
The ultrasound imaging technique has been demonstrated high reliability and a strong correlation for assessing spinal 
curvature when compared to conventional medical imaging methods such as X-ray.
Lv et al.~\cite{lv2020}  have assessed the Cobb angle using long-distance 3-dimensional ultrasound image systems. 
There was a high correlation ($r^{2}$ = 0.92) between the US and radiographic methods. Chen et al.~\cite{chen2016a} 
evaluated the Axial Vertebral Rotation (AVR) using the center of lamina (COL) method in both in-vitro and in-vivo studies. 
The COL method demonstrated strong agreement with Stokes' method in the in-vitro study. 
The Intraclass Correlation Coefficient (ICC) was between 0.84 and 0.85, 
and the Mean Absolute Difference (MAD) of 4.5$^\circ$ to 5.0$^\circ$. Zeng et al.~\cite{ge2020,zeng2021}  
automatically measured the Spinous Process Angle (SPA) by identifying spinous process curves using the gradient vector flow (GVF) snake method. 
The mean absolute differences (MADs) of SPAs obtained from the US and radiography were $3.4\pm2.4^\circ$ and $3.6\pm2.8^\circ$ for the two raters respectively.
Latest advancements in deep learning algorithms for scoliosis diagnosis have shown significant promise in improving 
the accuracy and efficiency of detection and segmentation tasks. 
Jiang et al.~\cite{jiang2022a} proposed the Ultrasound Global Guidance Block Network (UGBNet) to accurately segment 
and identify bony structures in spinal ultrasound images. The network incorporated a global guidance block module that 
merges spatial and channel attention mechanisms, allowing for learning long-range feature dependencies and contextual scale information. 
Zeng et al.~\cite{zeng2021} developed the stacked hourglass network (SHN) to accurately and quickly detect spinous process and lamina endpoints 
in ultrasound spine images. The SHN could effectively capture image features at all scales and the average processing time was just 10 minutes. 
 
 Although previous studies have applied ultrasound to assess the bony structures of the cervical spine~\cite{maartenvaneerd2014}, limited 
 research was explored to measure cervical spine curvature using 3D ultrasound.
 The objectives of this study are to utilize neural network algorithms to automatically segment bony 
 structures of the cervical spine and to propose an algorithm for measuring cervical 
 spine curvature as well as evaluating the feasibility of the proposed method. This article is organized as follows. Section.\ref{sec2} 
 introduces the principle of the nnU-Net network and the algorithms used for lamina curve identification. The method for data acquisition 
 and the assessment for lamina curvature angles will also be included. Section.\ref{sec3} illustrates the performance of the segmentation 
 and the measurement results of lamina curvature angles. The feasibility of the segmentation and the assessment of measurements will 
 be discussed in Section.\ref{sec4}. Section.\ref{sec5} will conclude the proposed work for this study.

\newpage

\section{Method}\label{sec2}
\begin{figure*}[ht]
	\centering
	\includegraphics[width=15cm]{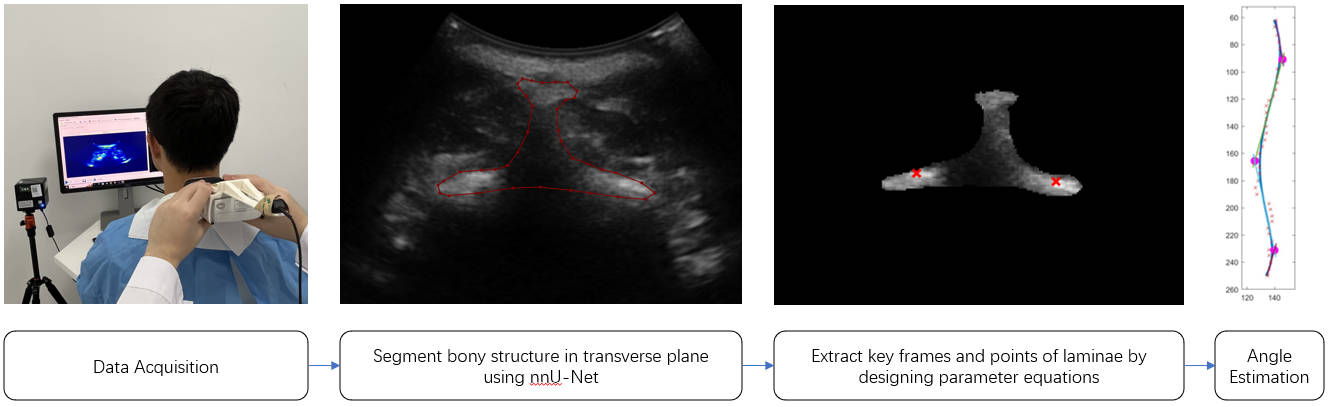}
    \caption{The flow chart of cervical lordosis measurement using 3D ultrasound imaging method
    }\label{Figure_flow}
\end{figure*}

The flowchart of this study is illustrated in Figure.\ref{Figure_flow}. 
The processing steps for cervical curvature detection were as follows: Firstly, 
3D ultrasound (US) data was acquired using a portable imaging system. 
Secondly, bony structures were segmented on the transverse ultrasound images. 
Thirdly, core points of laminae were extracted on the segmented images. 
Finally, the cervical lordosis, i.e., the cervical spine curve was automatically estimated based on these core points.

\subsection{Data Acquisition}
\begin{figure*}[b]
	\centering
	\includegraphics[width=15cm]{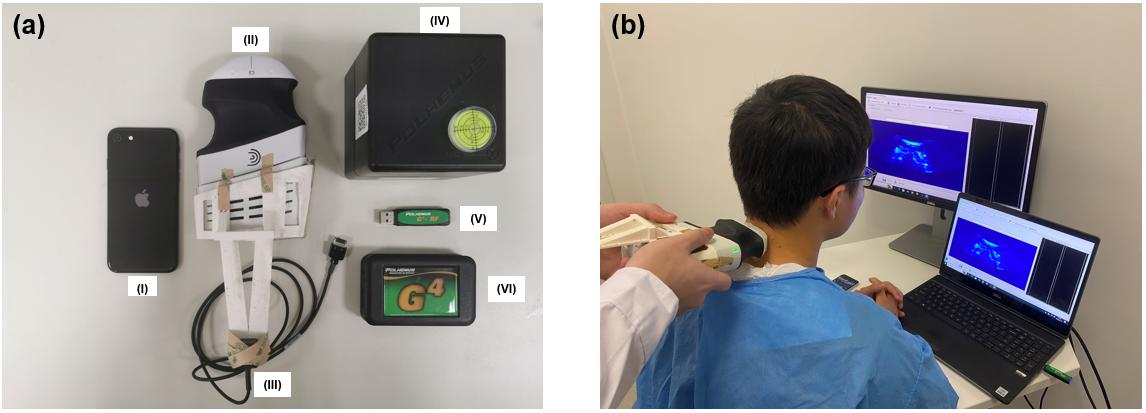}
    \caption{The 3D Cervical Spine US imaging system. (a) Image Acquisition System (I) iPhone SE (II) transportable US scanner. Tracking system. 
    (III) the 6 DOF sensor. (IV) the source. (V) RF/USB module. (VI) the
hub. (b) Scanning schematic}\label{Figure_Equip}
\end{figure*}

The equipments used for data acquisition included a Bluetooth and Wi-Fi enabled 2D ultrasound 
scanner (Clarius, C345-K, Canada) and a dual-mode tracking system (Polhemus, G4 unit, USA) 
were shown in Figure.\ref{Figure_Equip}a. Data processing and result display were handled by a 
host computer equipped with an Intel i7-8700K CPU @ 3.70 GHz and 32 GB RAM. 
The tracking system utilized a cube source (dimensions: 10.6 cm x 1.9 cm x 6.6 cm) 
to create an electromagnetic (EM) field. A 6 Degrees Of Freedom (6 DOF) location sensor 
was employed to obtain the US scanner's position and orientation information. 
A hub transmitted tracking data to the host computer wirelessly using an RF/USB module~\cite{chen2020}.

A total of 29 US scans were collected from 12 volunteers (10 males and 2 females). Three volunteers experienced cervical lordosis loss while 
the rest reported no neck pain or diseases. Ethics approval was granted from the local health ethics board and all participants signed consent 
forms prior to enrollment. 25 US scans were acquired in the neutral posture, which is similar to the standard posture for radiographic exams. 
Volunteers were seated with their backs straight and heads aligned forward at eye level. In addition, 4  ultrasound scans were acquired for the 
flexion posture to mimic the situation of cervical lordosis reversal. The volunteers were scanned from the bottom of the occipital bone to the seventh cervical vertebra. 
Each US scan consisted of 300-600 transverse frames. The resolution of each transverse frame was 640$\times$480.

\subsection{Segmentation Network Training} 

\begin{figure*}[b]
	\centering
	\includegraphics[width=15cm]{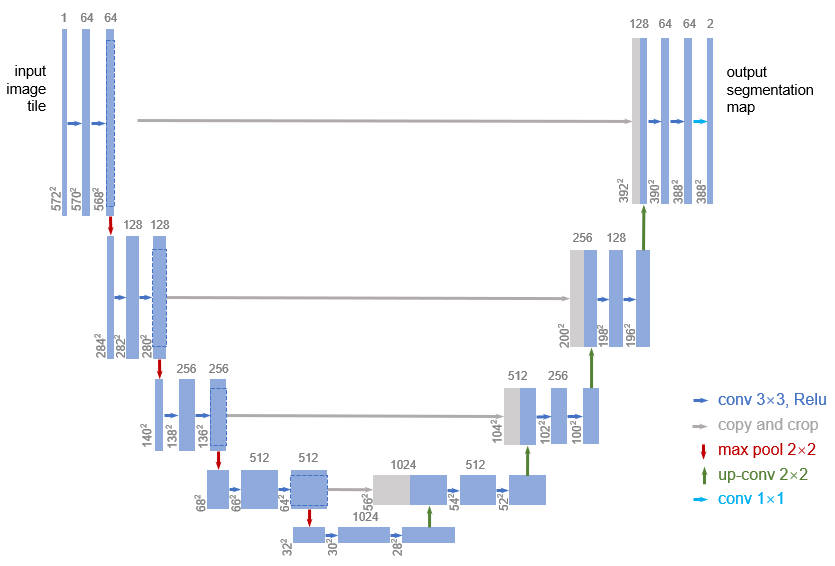}
    \caption{The network architecture of the nnU-Net.
    }\label{Figure_unet}
\end{figure*}

The deep learning algorithm for automatic segmentation of the cervical spine bony structures was trained using nnU-Net~\cite{isensee2021}, 
which is a self-configuring neural network designed specifically for biomedical image segmentation. Its adaptive mechanism analyzes the 
distinct properties of each dataset, thus can demonstrate better performance on medical image segmentation through customized preprocessing, 
network architecture, and training strategies. The structure example of the U-Net network is detailed in Figure.\ref{Figure_unet}.

In this study,  1,977 two-dimensional transverse ultrasound images from the cervical spine scans were used to train the nnU-Net network and 548 for testing. 
A 5-fold cross-validation was utilized to select the best-performing model during the training process. To achieve a balance between training speed and accuracy, 
the Poly learning rate scheduling strategy was used, starting with an initial rate of 0.01 and a decay exponent of 0.9 for gradual reduction. 
The nnU-Net network was trained on 500 epochs and evaluated using the Dice Similarity Coefficient (DSC). The training was conducted on a computer equipped 
with an NVIDIA RTX 2070 GPU with a memory capacity of 8 GB.

\subsection{Lamina Curve Identification and Angle Measurement}

After segmenting bony structures in each transverse image, the Fast Dot-Projection (FDP) method~\cite{chen2021}  was 
employed to reconstruct the 3D data volume. The identification and measurement of the lamina curve angle was divided into three phases:
\\[1ex]
\textit{Locating Key Frames in the Sagittal Planes} 

The key sagittal frames were obtained from the segmented 3D volume on both left and right sides, and on the frames,  the lamina's curves could be easily 
observed and accurately identified. Selection criteria were based on two image characteristics: the presence of the most complete bony structures with high image intensity. 
A parametric equation was developed to assess the significance of each frame:
 \begin{equation}
 Weight=\ln(\sum_{i=1}^{len } (\max(I_{row}(i))))\times len
\end{equation}
where \textit{len} represented the effective length of the bone area in the sagittal frame, and \textit{$I_{row}$} referred to an array of intensity values 
in the \textit{i}th row of the bone area. Frames with the maximum weights on the left and right sides were chosen as key frames. 
The sagittal frame containing the largest bony area did not always show the most complete structures, while the length of the high-intensity area indicated 
the length of the cervical spine, which was the main influencing factor of the curve measurement. Therefore, we assessed completeness based on the 
longitudinal length rather than the area to ensure maximum cervical spine length and clearest vertebral structure.
\\[1ex]
\textit{Identifying Core Points}

Intensity-weighted centroids in the lamina region were identified on each transverse frame, and the core points were then defined as the projection of 
these centroids on the key sagittal frames. 
\\[1ex]
\textit{Angle Measurement} 

In the last phase, the core points were firstly filtered using our DBSCAN algorithm~\cite{ge2020} in order to eliminate outliers. 
The 5$^{th}$ order polynomial was applied to fit all the core points into a lamina curve. The inflection points of the derived lamina curve were located.
 The tangent lines at all inflection points were subsequently solved. The lamina curve angles were then determined by computing the differences 
 in the angles between each pair of tangent lines at neighboring inflection points.

\subsection{In-vivo Evaluation}

The Dice Similarity Coefficient was calculated to evaluate the performance of the bony structure segmentation using nn-UNet. 
The ultrasound data collected from volunteers were utilized to validate the feasibility and reliability of cervical lordosis measurement using the proposed method.
The lamina curves passing through C2 to C7 derived on the two key sagittal frames were located and derived to indicate the cervical lordosis, 
and the curve angle measured on both left and right sides were then compared, i.e., the mean absolute difference (MAD), standard deviation (SD), 
and correlation (R) between the two angles of were computed to assess the reliability of the measurement results.

\section{Results}\label{sec3}
\begin{figure*}[t]
	\centering
	\includegraphics[width=16cm]{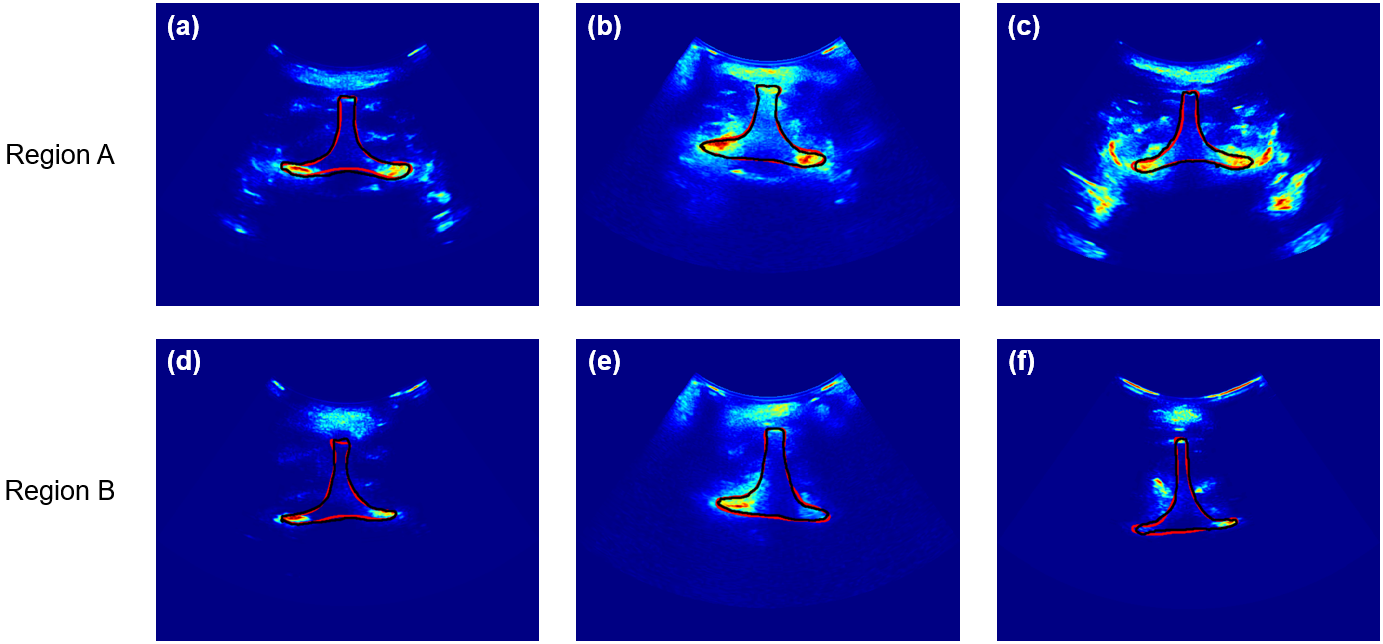}
    \caption{Bony structure predictions on various transverse ultrasound images. The black contour lines were ground truth, and the red contour lines were the 
    predictions from nnU-Net. (a)\&(b)\&(c) displayed examples from Region A, while (d)\&(e)\&(f) illustrated the examples from Region B. (b)\&(c)\&(e)\&(f) 
    showed the images with axial vertebral rotation; (e)\&(f) showed the images with the missing structure which contained one clear lamina and the other lamina unclear
    }\label{Figure_segresult}
\end{figure*}

The DSC value of the nnU-Net model in segmenting ROI was 0.973, indicating that the trained network showed high accuracy in segmenting the bony structures of the cervical spine. The cervical spine was divided into two regions, i.e., Region A including C2-C5, while Region B containing C6-C7.
Figure.\ref{Figure_segresult} showed the predicted results of transverse images from these two regions, 
including various rotation angles and the condition where one lamina was unclear.
The bony structures of Region B appeared more indistinct on ultrasound images than Region A due to uneven surfaces caused by less soft tissue and more protruding bones. 
The segmentation results demonstrated that our network could effectively recognize the differences among cervical vertebrae and accurately perform image segmentation.

\begin{figure*}[b]
	\centering
	\includegraphics[width=16cm]{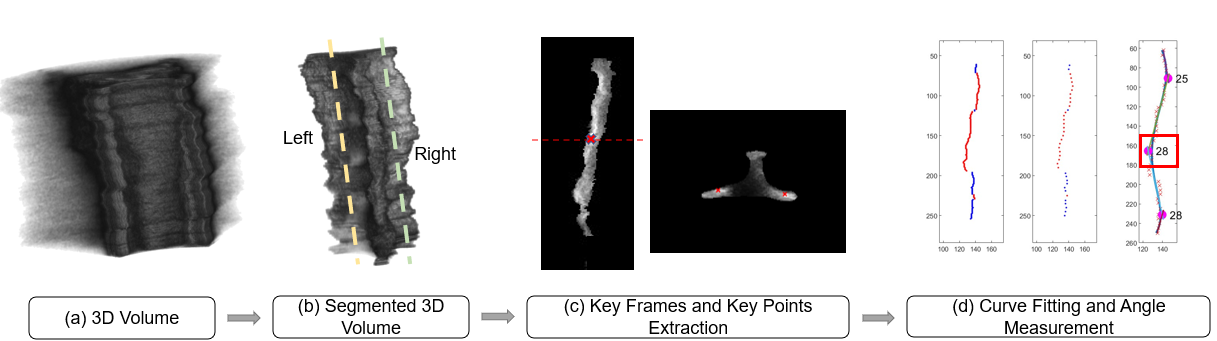}
    \caption{An illustration of cervical lordosis measurement. The left key frame was represented by a yellow dot line, while the right key frame was 
    represented by a green dot line in (b). (c) showed the core points as red dots shown. The lamina curve angle was displayed in Figure (d) as shown in the red box.
    }\label{Figure_result}
\end{figure*}

Figure.\ref{Figure_result} illustrated one example of the cervical lordosis measurement. 
It averagely took about five minutes to detect the bony structure using nnU-Net for each scan. 
After the 3-D reconstruction, the core points were highlighted on the ultrasound sagittal key frames as shown in Figure.\ref{Figure_result}(c). 
Twenty-two lamina curves were attained and automatically measured.

\begin{figure}[t]
    \centering
    \begin{minipage}{0.45\textwidth}
        \includegraphics[width=\linewidth]{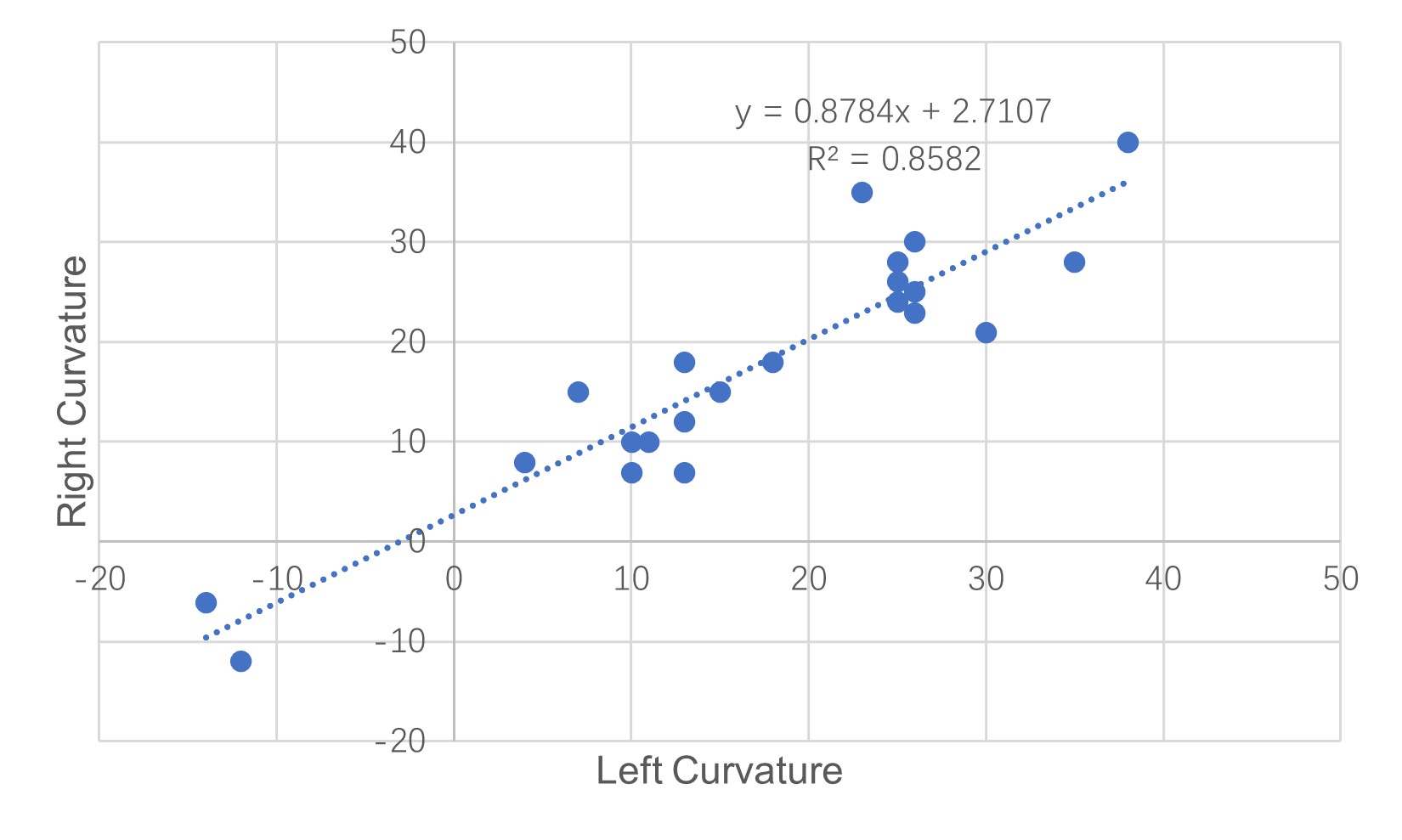} 
        \caption{The comparison of US Lamina Curve angles measured from left and right sides
    }\label{Figure_Chart}
    \end{minipage}
    \hfill 
    \begin{minipage}{0.45\textwidth}
        \includegraphics[width=\linewidth]{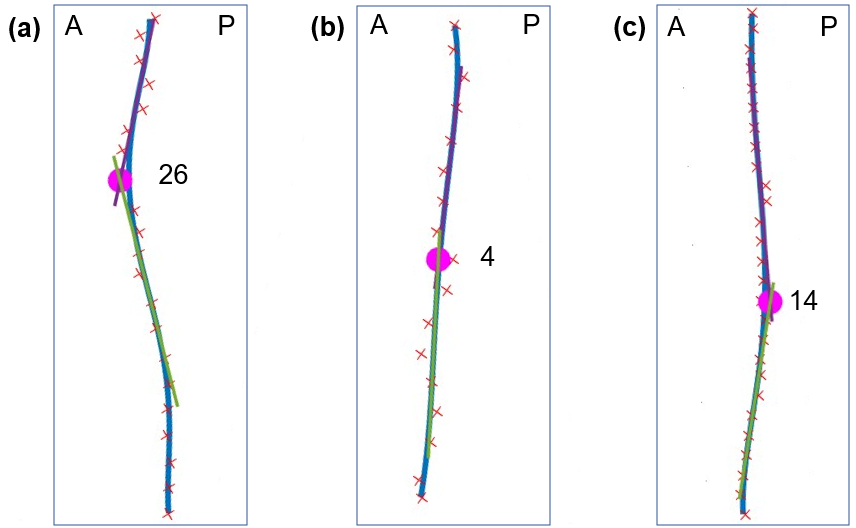}
        \caption{The comparison of  US Lamina Curve angles measured from three different status (a) Healthy subjects (b) Subjects experiencing loss of cervical 
        lordosis (c) Subjects mimicking the reversal of cervical lordosis
        }\label{Figure_measurementcompare}
    \end{minipage}
\end{figure}

Table.\ref{Table_Measurement} listed the detailed measurement data of all 22 lamina curve angles. In this study, 
the positive direction of cervical lordosis was defined as extension posture, while the negative direction corresponded to flexion posture. 
Typically, the cervical lordosis of a healthy subject in the neutral posture is in the positive direction. Figure.\ref{Figure_Chart} illustrated high agreement 
between left and right cervical lordosis. The $MAD\pm SD$and correlation for angles between two sides were $3.591 \pm 3.432^\circ$ and 0.926 respectively. 
The average measurement difference was less than the clinical acceptance error (5$^\circ$)~\cite{pruijs1994}. There were a total of 6 curves that showed large 
differences (>5$^\circ$) among all measurements.

Figure.\ref{Figure_measurementcompare} highlighted different lamina curve angles associated with three status: healthy subjects, 
subjects experiencing the loss of cervical lordosis and subjects mimicking the reversal of cervical lordosis. 
The illustration revealed notable differences between healthy and unhealthy subjects, which were evident not only in the shape of curves 
but also in the lamina curve angles. Therefore, our study indicated the feasibility of the cervical lordosis measurement, 
thus showing the potential of US technologies in the clinical diagnosis of cervical spine.

\begin{table}[t]
\centering
\caption{The measurement results of overall 22 US Lamina Curves}
\label{Table_Measurement}
\resizebox{\textwidth}{!}{%
\begin{tabular}{ccccc}
\Xhline{5\arrayrulewidth}
Status &
  Data NO. &
  Left($^\circ$) &
  Right($^\circ$) &
  \begin{tabular}[c]{@{}c@{}}Absolute \\ Difference($^\circ$)\end{tabular} \\ \hline
\multirow{13}{*}{Healthy subjects} &
  HS001 &
  25 &
  28 &
  3 \\
 & HS002  & 7   & 15             & 8     \\
 & HS003  & 13  & 12             & 1     \\
 & HS004  & 23  & 35             & 12    \\
 & HS005  & 26  & 23             & 3     \\
 & HS006  & 26  & 30             & 4     \\
 & HS007  & 38  & 40             & 2     \\
 & HS008  & 15  & 15             & 0     \\
 & HS009  & 13  & 18             & 5     \\
 & HS010  & 30  & 21             & 9     \\
 & HS011  & 35  & 28             & 7     \\
 & HS012  & 26  & 25             & 1     \\
 & HS013  & 25  & 24             & 1     \\ \hline
\multirow{4}{*}{\begin{tabular}[c]{@{}c@{}}Subjects experiencing the \\ loss of cervical lordosis\end{tabular}} &
  LCS001 &
  11 &
  10 &
  1 \\
 & LCS002 & 4   & 8              & 4     \\
 & LCS003 & 10  & 10             & 0     \\
 & LCS004 & 13  & 7              & 6     \\ \hline
\multirow{5}{*}{\begin{tabular}[c]{@{}c@{}}Subjects mimicking the \\ reversal of cervical lordosis\end{tabular}} &
  MCS001 &
  25 &
  26 &
  1 \\
 & MCS002 & 18  & 18             & 0     \\
 & MCS003 & 10  & 7              & 3     \\
 & MCS004 & -12 & -12            & 0     \\
 & MCS005 & -14 & -6             & 8     \\ \hline
 &        &     & MAD($^\circ$)  & 3.591 \\
 &        &     & SD($^\circ$)   & 3.432 \\
 &        &     & R & 0.926 \\ 
\Xhline{5\arrayrulewidth}
\end{tabular}%
}
\end{table}

\section{Discussion}\label{sec4}
\subsection{Feasibility Analysis}
\begin{figure*}[b]
	\centering
	\includegraphics[width=13cm]{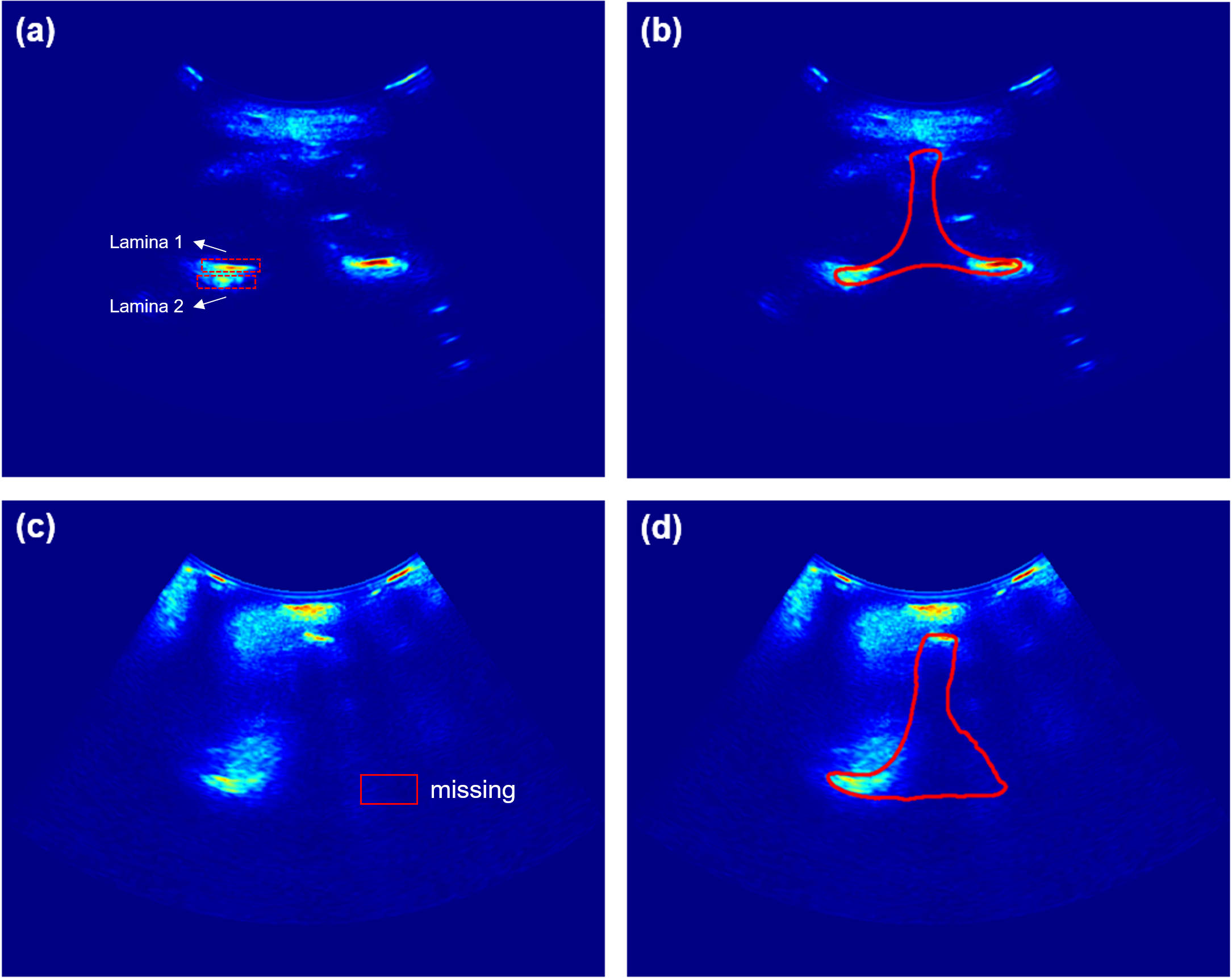}
    \caption{Two cases of bad segmentation results (a)(c) Ground Truth, (a) The red box outlined two adjacent laminae (b) Inaccurate differentiation 
    of correct lamina pairs (c) The red box represented the lamina region that should be segmented (d) Improper depicting of precise bony structures 
    with missing pairs of laminae.
    }\label{Figure_dis1}
\end{figure*}

The nnU-Net network acquired high performance in segmenting cervical bony structures. Generally, lamina features are apparent because they are 
typically located in the central region of transverse images with high intensity. Additionally, due to the anatomical structure of the vertebra, 
there is a strong correlation between spinous processes and laminae. The network could utilize this correlation to predict bone structures whose 
shape is close to a triangle. However, the small number of training datasets caused the overfitting problems. As illustrated in Figure.\ref{Figure_dis1}a, 
when the ultrasound probe scanned the junction between adjacent cervical vertebrae, it simultaneously received reflected signals from the two vertebrae 
since the thick ultrasound probe caused a thick scanning slice in the axial direction covering both vertebrae. 
Consequently, the ultrasound image displayed two adjacent laminae, leading to errors in the segmentation process. Figure.\ref{Figure_dis1}b 
showed inaccurate segmentation results when one lamina was missing in the ultrasound image. The absence of adequate brightness information at 
certain bone locations caused the network to erroneously assign incorrect weights to noisy signals when attempting to force out a segmentation result. 
This misallocation would lead to abnormal-shaped segmented regions. Nevertheless, the challenges from segmentation showed minimal impact when measuring 
laminae curve angles since the DBSCAN method could filter outliers and reduce the influence of the uncertainties caused by incorrect segmentation. 
In future research, more data from diverse subjects will be collected to enhance the network’s generalizability.

This study did not incorporate clinical data due to the limited access to the clinic patients. Therefore, in-vivo volunteer data were utilized to 
validate the reliability of the proposed method. The consistency of the curve measurement results on both the left and right sides demonstrated a 
reliable outcome of the ultrasound cervical imaging method. Additionally, a significant difference in measurement angles was clearly observed 
in subjects under different conditions, such as with the loss of cervical lordosis, healthy status in the neutral posture, or the flexion posture 
with intending forward bending. However, there were some cases showing a large discrepancy over 5$^\circ$ between the left and right sides as 
depicted in Table.\ref{Table_Measurement}. This discrepancy mainly arose from non-standardized postures during the scanning process.  
The slight head twist to the side would cause
considerable cervical vertebral rotation which can make one side higher than the other side of the neck and then 
lead to different cervical curves on different sagittal frames. Excluding this exceptional case, the $MAD\pm SD$ between the left and right 
sides would decrease to $3.190\pm 2.943^\circ$, and a more robust correlation (R=0.946) between the two sides.

\subsection{Difficulty in Distinguishing Individual Cervical Vertebrae}

For ultrasound spine imaging of scoliosis research, individual vertebra could be easily identified and accurately annotated~\cite{tang2022}. 
However, distinguishing individual cervical vertebra in the posterior images was a significant challenge in this study. 
The spaces between cervical vertebrae are narrow, and part of the vertebrae are overlapped~\cite{patwardhan2018}. Consequently, 
when the ultrasound probe traversed the gaps between the cervical vertebrae, it frequently captured adjacent vertebrae at the same time. 
Therefore, it was difficult to clearly distinguish the spaces between the cervical vertebrae on the US  sagittal images. 
Future work will focus on developing imaging processing methods that can differentiate individual cervical vertebra.

\subsection{C1 Identification}
The vertebra of C1 was often obscured by the skull during ultrasound scanning due to its proximity to the inferior border of the occipital bone. 
Therefore, it was difficult to identify C1 on the US images. Additionally, C1 has a significant anatomical difference due to the absence of laminae compared to C2-C7. 
As a result, C1 was not included in the lamina curve calculations in this study. Moreover, gold standard methods of cervical lordosis measurements, 
such as the Cobb angle, also exclude C1 due to its minimal impact on overall cervical lordosis. 
Thus, the challenge of imaging C1 did not affect the accuracy of the curvature measurements.

\section{Conclusion}\label{sec5}
In this study, an automatic method based on nnU-Net was proposed for segmenting bony structures in cervical spine ultrasound images, 
and could accurately segment bony structures with DSC of 0.973. The cervical spine curve was derived and the curvature was precisely 
measured based on the extracted features on the key frames from the segmented 3D Volumes. The results showed a high agreement of the 
laminae curve angles measured on the left and right sides of the cervical spine with the  $MAD\pm SD$  were $3.591\pm3.432^\circ$ and 
the correlation of 0.926 indicating a reliable measurement result. In addition, the ultrasound cervical images exhibited significant differences 
for the cases with different conditions of cervical lordosis such as loss of cervical lordosis, neutral posture, and flexion posture. 
Our proposed method showed the potential of diagnosing the loss of cervical lordosis using the 3D ultrasound cervical imaging technique. 
In future work, more clinical data will be acquired to improve the performance and generalization of the algorithm. 
Comparative studies between ultrasound imaging and other modalities such as X-ray and MRI will be implemented to evaluate the accuracy 
and effectiveness of the proposed method.

\section*{Acknowledgement}
The authors are profusely grateful for the sponsorship from the Natural Science Foundation of China (NSFC)
under Grant No.12074258.
And the authors also appreciated the support from the Platform of Mechatronics, Energy, and Electronic Devices.
\section*{References}

\bibliographystyle{ieeetr}
\bibliography{JPCSLaTeXGuidelines}
\end{document}